\documentclass[oupdraft,a4paper]{biott}
\usepackage[colorlinks=true, urlcolor=citecolor, linkcolor=citecolor, citecolor=citecolor]{hyperref}

\usepackage{float}
\usepackage{amsmath,amssymb,mathrsfs,graphicx,bbm}
\usepackage{amsthm}

\begin{document}
\title{Adding experimental treatment arms to Multi-Arm Multi-Stage platform trials in progress}

\author{THOMAS BURNETT\\[-5pt]
\textit{MRC Biostatistics Unit, University of Cambridge, Cambridge, CB2 0SR}
\\[2pt]
FRANZ K\"ONIG$^\ast$\\[-5pt]
\textit{Section for Medical Statistics,CeMSIIS,Medical University of Vienna, Vienna 1090,Austria}
\\
{franz.koenig@meduniwien.ac.at}
\\[2pt]
THOMAS JAKI\\[-5pt]
\textit{Department of Mathematics and Statistics, Lancaster University, Lancaster LA1 4YF, UK\\
MRC Biostatistics Unit, University of Cambridge, Cambridge, CB2 0SR}
}

\markboth%
{T. Burnett and others}
{Adding treatment arms to trials in progress}

\maketitle

\footnotetext{To whom correspondence should be addressed.}

\begin{abstract}
{Multi-Arm Multi-Stage (MAMS) platform trials are an efficient tool for the comparison of several treatments.  Suppose we wish to add a treatment to a trial already in progress, to access the benefits of a MAMS design. How should this be done?

The MAMS framework requires pre-planned options for how the trial proceeds at each stage in order to control the family-wise error rate.  Thus, it is difficult to make both planned and unplanned design modifications.  The conditional error approach is a tool that allows unplanned design modifications while maintaining the overall error rate.  In this work, we use the conditional error approach to allow adding new arms to a MAMS trial in progress.

We demonstrate the principles of incorporating additional hypotheses into the testing structure.  Using this framework, we show how to update the testing procedure for a MAMS trial in progress to incorporate additional treatment arms.  Simulations illustrate the possible operating characteristics of such procedures using a fixed rule for how and when the design modification is made.}
{multi-arm multi-stage (MAMS), adaptive designs, conditional error, design modification}
\end{abstract}

\section{Introduction}

During Phase $\mathrm{II}$ of the drug development process it is common to have several competing treatments, these may be different doses of the same drug or entirely different treatment regimes. \cite{jaki2016designing} note that, given the high failure rate and cost of Phase $\mathrm{III}$ trials, it is key to that careful consideration be given to which treatments should be carried forward for further study. Multi-arm multi-stage trials (MAMS) \citep{royston2003novel,jaki2013considerations,wason2012optimal} compare several experimental treatments with a common control allowing for the efficient selection of appropriate treatments \citep{jaki2015multi}.\\

MAMS trials reduce the expected number of patients by dropping treatments that are demonstrated to be ineffective/showing lack of promise or stopping the trial altogether if efficacy has been demonstrated. Given the multiple hypotheses and highly adaptive nature of the design, MAMS studies require specialist testing methodology in order to control the error rate of the trial \citep{stallard2003sequential}. \cite{magirr2012generalized} introduced the generalised Dunnett family of tests, where group sequential testing boundaries are defined to account for the multiple analyses, while accounting for the correlation introduced by the comparison of several experimental arms to a common control \citep{dunnett1955multiple}; \cite{urach2016multi} extend this directly defining all elements of the testing procedure. Alternatively fully flexible testing methods have been proposed (for example, \citep{bretz2006confirmatory,schmidli2006confirmatory,posch2005testing,koenig2008adaptive,bauer1999combining}), allowing decisions about which arms should remain in the study to function separately from the hypothesis testing. Both methods require the pre-definition of all study hypotheses, so that the overall testing procedure may be constructed to give strong control of the Family-Wise Error Rate (FWER) \citep{dmitrienko2009multiple}.\\ 

It is possible that not all experimental treatments are available at the start of the trial as, for example, see in the STAMPEDE trial \citep{sydes2009issues}. STAMPEDE started with five comparisons and subsequently added several more to the protocol. Including further experimental treatments into the trial in progress maintains the benefits of a MAMS design reducing logistical and administrative effort, speeding up the overall development process \citep{parmar2008speeding}, efficiency in the multiple comparisons and allowing direct comparisons of the treatments within the same trial.\\

Treatments may be added to the trial in progress by adjusting the pre-planned testing structure provided no use has been made of the data observed in the trial (thus requiring no interim analysis has been conducted). \cite{bennett2020designs} demonstrate how to suitably adjust the sample size for each treatment arm for such additions. It is possible that treatments may become available after some interim analysis, our methods allow modification at any stage of the trial with the only restriction being that no conclusion of statistical significance has been made. \\

The conditional error approach \citep{proschan1995designed} allows for design modifications during the course of a trial, where these modifications have not been pre-planned. It has been shown these modifications may be accounted for in the setting of treatment selection \citep{koenig2008adaptive,magirr2014flexible} however, adding hypotheses to a testing framework requires further restrictions on any introduced hypotheses \citep{hommel2001adaptive}. We propose a general framework using these principles for the inclusion of additional hypotheses to a testing procedure that allows the inclusion of existing trial information. We show how to apply this in the setting of MAMS designs, demonstrating how to construct an appropriate hypothesis testing structure for the updated trial such that the FWER is strongly controlled. 

\section{Altering a trial in progress}

\subsection{A two arm trial}
\label{sec:twoarm}
Suppose we  plan a two arm trial with a continuous outcome to compare a new treatment, $T_1$, and a control, $T_0$. Let $\mu_1$ and $\mu_0$ be the expected responses for patients on treatments $T_1$ and $T_0$ respectively, and define the treatment effect as $\theta_1 = \mu_1 - \mu_0$. We investigate the one sided null hypothesis $H_{01}:\theta_1 \leq 0$.\\

The trial will recruit a total of $n$ patients randomised equally between treatment and control. Let $X_{i,k} \sim N(\mu_k,\sigma^2)$ for $i=1,...,n/2$ and $k=0,1$ then $\hat{\theta}_1$ is the estimate of the treatment effect. For $\xi_1 = \frac{\theta_1\sqrt{n}}{2\sigma}$ this has corresponding Z-value,
$$Z_1 = \frac{\hat{\theta}_1\sqrt{n}}{2\sigma}\sim N(\xi_1,1).$$
We reject $H_{01}$ at leve $\alpha$ when $Z_1 > \Phi^{-1}(1-\alpha)$, where $\Phi$ is the standard normal cdf.

\subsection{Adding a treatment}
\label{sec:simpadd}

Suppose for $\tau \in (0,1)$ after $\tau n$ observations a new treatment, $T_2$, becomes available. Let $\mu_2$ be the expected response for patients receiving this new treatment and define the corresponding treatment effect by $\theta_2 = \mu_2 - \mu_0$ with corresponding null hypothesis $H_{02}:\theta_2 \leq 0$. \\

Suppose, we maintain the pre-planned elements of the trial concerning treatments $T_1$ and $T_0$, such as the same sample size per treatment. Notationally it is convenient to define stage 1 and stage 2 consisting of the patients recruited before and after the treatment is added. From the stage 1 data we find
$$Z_1^{(1)} \sim N(\xi_1\sqrt{\tau},1)$$
and from the stage 2 data we find
$$Z_1^{(2)} \sim N(\xi_1\sqrt{1-\tau},1).$$
The overall Z-value may be reconstructed from the stagewise Z-values
$$Z_1 = \sqrt{\tau}Z_1^{(1)} + \sqrt{1-\tau}Z_1^{(2)}.$$\\

We recruit a further $(1-\tau)n/2$ patients to $T_2$ in stage 2, maintaining equal randomisation to all treatments. Since $T_2$ is added to the trial for stage 2 for $\xi_2 = \frac{\theta_2\sqrt{n}}{2\sigma}$
$$Z_2 \sim N(\xi_2\sqrt{1-\tau},1),$$
is based only on the data available from the second stage of the trial from which we construct the Z-value. Due to the common control and equal randomisation $Z_1^{(2)}$ and $Z_2$ have correlation $1/2$.

\subsection{Hypothesis testing}
\label{sec:addtwo}

For the two arm trial we constructed our hypothesis test in order to control the type $\mathrm{I}$ error rate at some pre-defined level $\alpha$. A natural extension in the case of multiple hypotheses is the Family-Wise Error rate (FWER), for the event $R$ that we reject one or more true null hypothesis the FWER is defined as $P_{\boldsymbol{\theta}}(R)$.\\

\begin{figure}[ht]
\makebox[\linewidth]{
\includegraphics[scale=0.5]{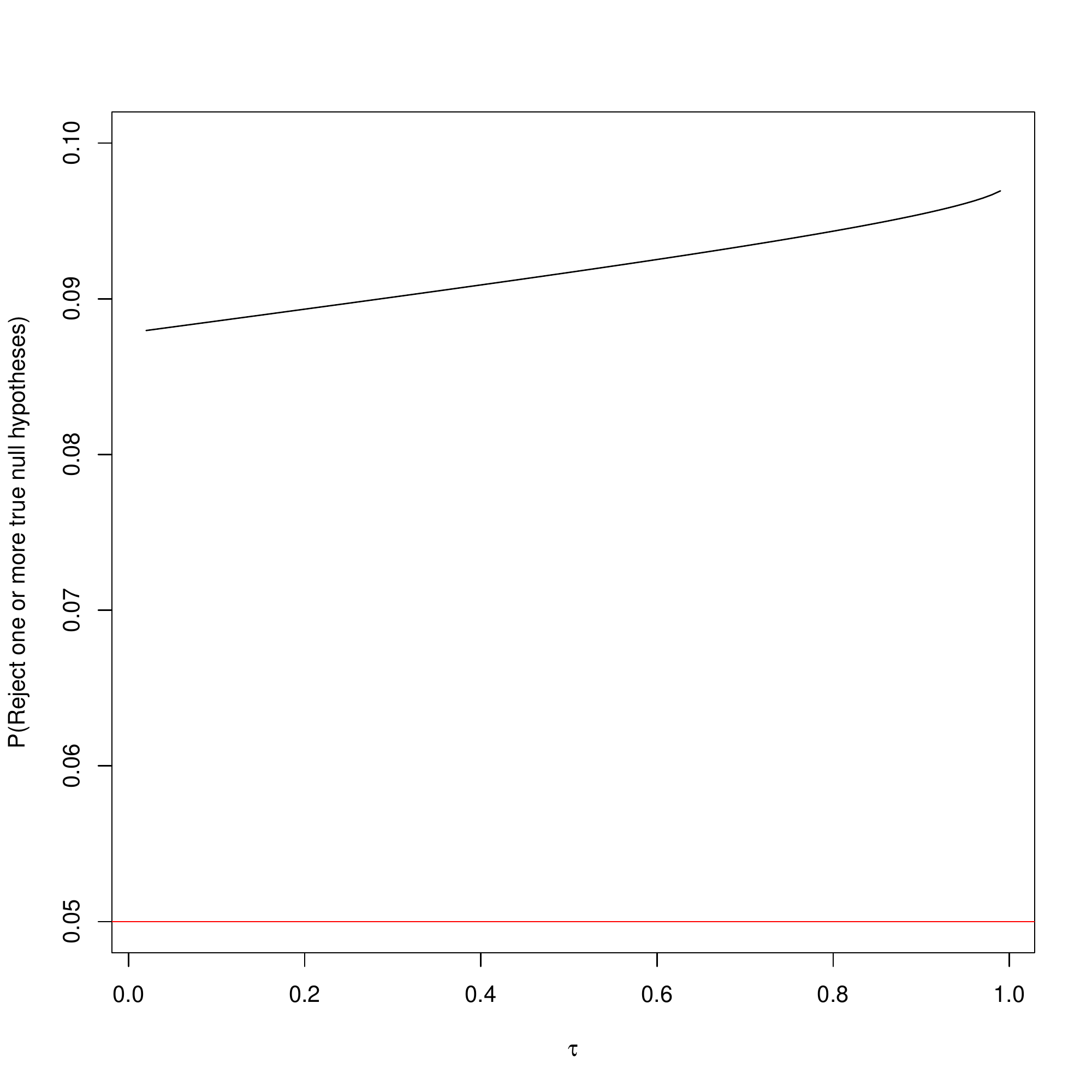}}
\label{fig:FWERplot}
\caption{Inflation in the FWER when an additional hypothesis is added to an ongoing two arm trial. The red line is the nominal FWER $\alpha = 0.05$ as per the design and the black line is the actual FWER for the given $\tau$.}
\end{figure}

Suppose, when adding $T_2$ we test each null hypotheses at a nominal level $\alpha = 0.05$, Figure~\ref{fig:FWERplot} shows the impact on the FWER as we vary $\tau$. Under extremes of $\tau=0$ and $\tau=1$ the trial is not altered and should be designed accordingly to achieve a FWER of $\alpha$. For all values in between we see that the FWER is inflated when compared to the nominal $\alpha$.\\

As is typical in a confirmatory setting \citep{dmitrienko2009multiple} we require strong control of the FWER, that is
\begin{equation}
P_{\boldsymbol{\theta}}(R) \leq \alpha \text{ for all }\boldsymbol{\theta}=(\theta_1,\theta_2).
\label{eq:FWER}
\end{equation}
\cite{sugitani2018flexible} propose methods that account for the introduction of the additional hypothesis, testing any introduced hypothesis based strictly on the data collected after their introduction at level $\alpha$ \citep{hommel2001adaptive}. We build on this approach, incorporating existing information where possible.\\

We construct an overall closed testing procedure \citep{marcus1976closed} that accounts for the adaptive nature of the trial within each test \citep{koenig2008adaptive}. This requires tests of $H_{01}$, $H_{02}$ and $H_{0,12} = H_{01} \cap H_{02}: \theta_1 \leq 0 \text{ } \& \text{ } \theta_2 \leq 0$. Rejecting $H_{01}$ globally when the local level $\alpha$ tests of $H_{01}$ and $H_{0,12}$ are rejected and $H_{02}$ globally when the local level $\alpha$ tests of $H_{01}$ and $H_{0,12}$ are rejected.\\

No changes have been made to the recruitment or analysis concerning $H_{01}$, so as before we reject $H_{01}$ when $Z_1 > \Phi^{-1}(1-\alpha)$ at the end of the trial. It is useful to discuss constructing this test using the conditional error principle \citep{proschan1995designed}.  Given $z_1^{(1)}$ we define the conditional error rate
$$A(z_1^{(1)}) = \mathbbm{P}_{\theta_1 = 0}(\text{Reject }H_{01}|Z_1^{(1)}=z_1^{(1)}).$$
The probability of rejecting the null hypothesis for the remainder of the trial must not exceed $A(z_1^{(1)})$. Writing the test in terms of the stage 2 observations while incorporating the stage 1 data, we reject $H_{01}$ when $Z_1^{(2)} > \Phi^{-1}(1-A(z_1^{(1)}))$. Let $f(z_1^{(1)})$ be the probability density function of $z_1^{(1)}$, under $H_{01}$
\begin{equation}
\int_{z_1^{(1)}}f(z_1^{(1)})A(z_1^{(1)})\mathrm{d}z_1^{(1)} = \alpha,
\label{eq:con}
\end{equation}
which in turn guarantees control of the error rate at the pre-specified level $\alpha$, that is
$$\mathbbm{P}_{\theta_1 = 0}(\text{Reject }H_{01}) = \alpha.$$\\

There is no existing for $H_{02}$ and thus the test must be constructed purely based only on the stage 2 trial data used to construct $Z_2$. We reject the test for $H_{02}$ when $Z_2 > \Phi^{-1}(1-\alpha)$.\\

There is no pre-planned test for  $H_{0,12}$, however there is pre-existing information for $H_{01}$ in the form of $Z_1^{(1)}$. \cite{hommel2001adaptive} show how to use such first stage information in the test of an intersection hypothesis, when adding some initially excluded hypotheses after an interim analysis which we apply to the added hypothesis. Clearly if $H_{0,12}$ is true this implies that $H_{01}$ is also true. Since $H_{01}$ is true we compute the conditional error rate $A(z_1^{(1)})$ as described previously, furthermore under $H_{01}$, $z_1^{(1)}$ is distributed such that equation~\ref{eq:con} holds as before. Thus we may construct the test of $H_{0,12}$ at the end of the trial at level $A(z_1^{(1)})$ allowing for the incorporation of the stage one data given by $Z_1^{(1)}$.\\

For example, consider a Dunnett test \citep{dunnett1955multiple} for $H_{0,12}$. Let
$$Z_D = max(Z_1^{(2)},Z_2)$$
and define the distribution
$$\
\begin{pmatrix}
X \\ Y
\end{pmatrix}
 \sim N\left(
\begin{pmatrix}
0 \\ 0
\end{pmatrix}
,
\begin{pmatrix}
1 & 1/2\\1/2 & 1
\end{pmatrix}\right).$$
We construct the Dunnett p-value,
$$P_D = \mathbbm{P}(X>Z_D \cup Y>Z_D)$$
and may reject $H_{0,12}$ when $P_D < A(Z_1^{(1)})$.\\

The choice to recruit a further $(1-\tau)n/2$ patients to each treatment after the interim analysis is not required. The total number of patients recruited in stage 2 is free to vary however, changes to the ratio of patients on treatment and control requires slight further modification (although the ratio remains fixed after the modification is made). If the ratio of patients between the existing treatment and control differ before and after the design modification it is no longer possible to weight the Z-values in order to recover the pooled test statistic, in which case $Z_1$ would need to be constructed by using the weighted inverse normal \citep{bauer1994evaluation, lehmacher1999adaptive, hartung1999note} with weights defined at the time the modification is made. 

\subsection{Simulation study}
\label{sec:sim}

For combincations $(\xi_1,\xi_2)$, with $\sigma/n =1$, $\delta = \Phi^{-1}(0.95)+\Phi^{-1}(0.9)$ and $\tau =0.5$ we simulate 1,000,000 realisations of $Z_1$ and $Z_2$ assuming equal sample size in each treatment at each stage in R \citep{R}. Table~\ref{tab:tests} shows estimates of the probabilities of an error for the local hypothesis tests, as required this is $\alpha$ whichever combination of null hypotheses are true.\\

\begin{table}[ht]
\makebox[\linewidth]{
\begin{tabular}{c c|c c c c}
$\xi_1$ & $\xi_2$ & $P(\text{Reject test of }H_{01})$ & $P(\text{Reject test of }H_{02})$ & $P(\text{Reject test of }H_{0,12})$ & FWER\\\hline
0 & 0 & \textbf{0.05} & \textbf{0.05} & \textbf{0.05} & 0.05\\
$\delta$ & 0 & 0.90 & \textbf{0.05} & 0.86 & 0.05\\
0 & $\delta$ & \textbf{0.05} & 0.66 & 0.36 & 0.05
\end{tabular}}
\caption{Probabilities of rejecting components of the closed testing procedure under proposed testing procedure, type $I$ errors highlighted in bold, $\delta = \Phi^{-1}(0.95)+\Phi^{-1}(0.9)$ such that we have power of $0.9$ when testing $H_{01}$ in the original trial.}
\label{tab:tests}
\end{table}

We compare the overall trial performance of the method proposed above with basing the test for the intersection hypothesis only on evidence for $H_{01}$, that is we reject $H_{0,12}$ when $Z_1 > \Phi^{-1}(1-\alpha)$ (treating the first null hypothesis as a gate keeping procedure \citep{dmitrienko2007gatekeeping}). In both procedures $Z_1^{(1)}$ is used in the test of $H_{0,12}$ by the argument that $H_{0,12} \implies H_{01}$. Table~\ref{tab:probs} shows the global rejection probabilities for the null hypotheses for each testing method. Both testing procedures gives strong control of the FWER. The probabilities of rejecting false $H_{01}$ do not differ largely, with an increase of at most 0.04 for the gate keeping procedure.  When both null hypotheses are false, there is a small decrease of 0.03 in the probability of rejecting $H_{02}$ for the gate keeping procedure. When $H_{01}$ is true and $H_{02}$ is false the gate keeping procedure cannot reject $H_{02}$ without making an error in rejecting $H_{01}$ and thus our proposed procedure increases the probability of rejecting the $H_{02}$ by 0.29. The small advantage for testing $H_{01}$ for the gate keeping procedure is outweighed by the ability to reject $H_{02}$ when there is a low probability of rejecting $H_{01}$ when taking an integrated approach to the test of the intersection hypothesis.

\begin{table}[ht]
\makebox[\linewidth]{
\begin{tabular}{c c|c c c c}
\multicolumn{6}{c}{Dunnett procedure for testing the intersection hypothesis}\\
$\xi_1$ & $\xi_2$ & $P(\text{Globally reject }H_{01}\text{ only})$ & $P(\text{Globally reject }H_{02}\text{ only})$ & $P(\text{Globally reject both})$ & $P(\text{Globally reject any})$ \\\hline
0 & 0 & \textbf{0.03} & \textbf{0.01} & \textbf{0.01} & \textbf{0.05}\\
$\delta$ & 0 & 0.81 & \textbf{0.00} & \textbf{0.05} & 0.86\\
0 & $\delta$ & \textbf{0.00} & 0.29 & \textbf{0.04} & 0.33\\
$\delta$ & $\delta$ & 0.26 & 0.03 & 0.62 & 0.91\\
\multicolumn{6}{c}{}\\
\multicolumn{6}{c}{Gate keeping procedure for testing the intersection hypothesis}\\
$\xi_1$ & $\xi_2$ & $P(\text{Globally reject }H_{01}\text{ only})$ & $P(\text{Globally reject }H_{02}\text{ only})$ & $P(\text{Globally reject both})$ & $P(\text{Globally reject any})$ \\\hline
0 & 0 & \textbf{0.04} & NA & \textbf{0.01} & \textbf{0.05} \\
$\delta$ & 0  & 0.85 & NA & \textbf{0.05} & 0.90\\
0 & $\delta$ & \textbf{0.01} & NA & \textbf{0.04} & 0.05\\
$\delta$ & $\delta$ & 0.28 & NA & 0.62 & 0.90
\end{tabular}}
\caption{Probabilities of global rejection of null hypothesis using the conditional error approach, type $I$ errors highlighted in bold, $\delta = \Phi^{-1}(0.95)+\Phi^{-1}(0.9)$ such that we have power of $0.9$ when testing $H_{01}$ in the original trial.}
\label{tab:probs}
\end{table}

\begin{figure}[ht]
\makebox[\linewidth]{
\includegraphics[scale=0.65]{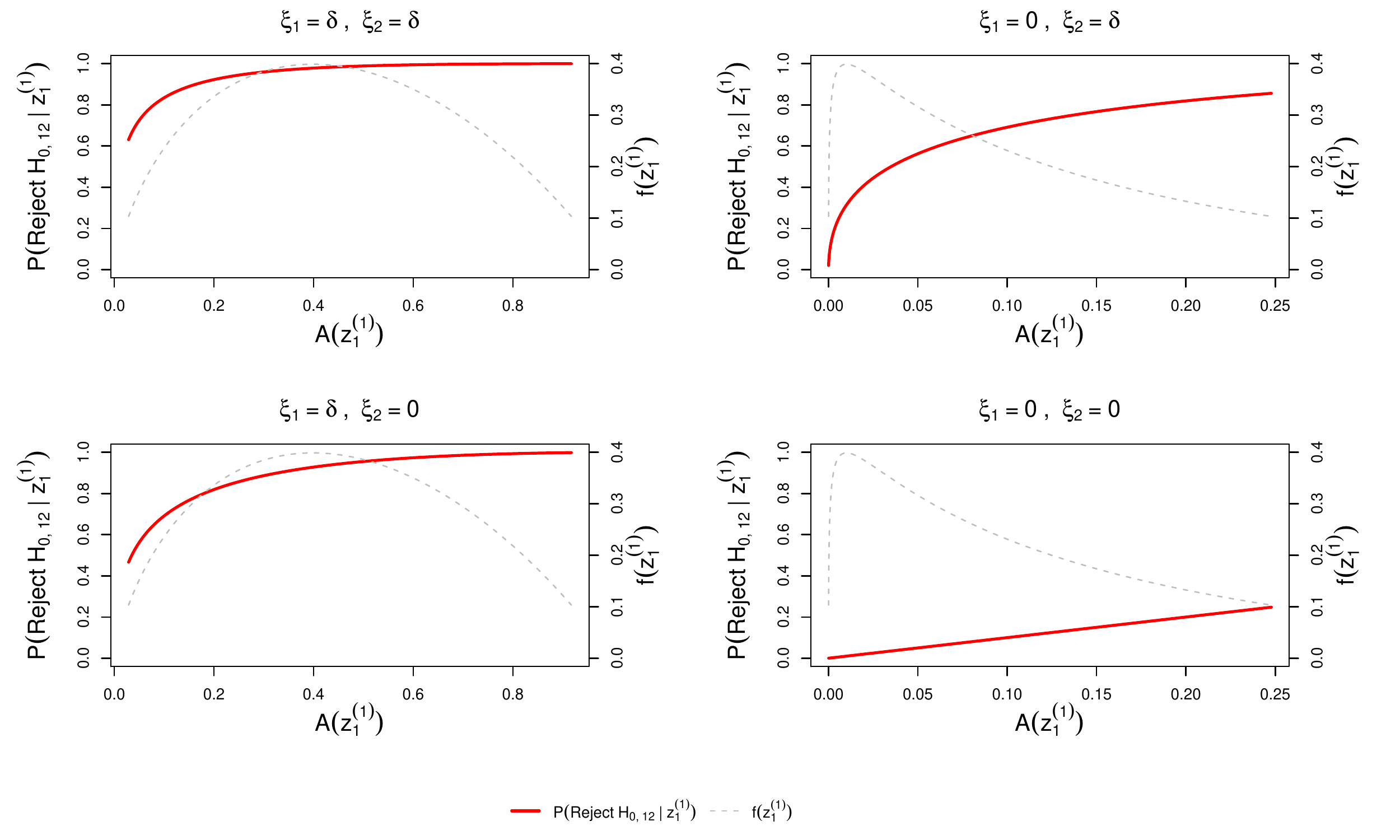}}
\caption{Conditional error rate, $A(z_1{(1)})$, against probability of rejecting the intersection hypothesis $P(\text{Reject }H_{0,12}|z_1^{(1)})$ and corresponding density of conditional error $f(z_1^{(1)}$, $\delta = \Phi^{-1}(0.95)+\Phi^{-1}(0.9)$ such that we have power of $0.9$ when testing $H_{01}$ in the original trial.}
\label{fig:condpow}
\end{figure}

In Figure~\ref{fig:condpow} we examine the probabilities of rejecting the intersection hypothesis $H_{0,12}$ for all combinations of $H_{01}$ and $H_{02}$ true and false. When $H_{01}$ is false the conditional error is likely to be higher than the pre-planned $\alpha$, giving a high chance of rejecting $H_{0,12}$; when $H_{01}$ is true and $H_{02}$ is false there is a small reduction in the probability of rejecting $H_{0,12}$, this explains the deficit of our proposed procedure when $\xi_1 = \delta$ and $\xi_2 = 0$. Conversely when $H_{01}$ is true the conditional error is likely to be quite low: when both null hypotheses are true this corresponds to a low probability of rejecting $H_{0,12}$ however, when $H_{02}$ is false we recover some possibility of rejecting $H_{0,12}$ allowing us to reject $H_{02}$ globally.

\section{General rule for adding hypotheses}
\label{sec:gen}

Suppose there is a set of existing null hypotheses $E$ with a pre-planned closed testing procedure, and we wish to add a set of new null hypotheses $N$. Let $H_e$ be the intersection of some subset of the existing null hypotheses $e \subseteq E$ and $H_n$ be the intersection of some subset of the new hypotheses $n \subseteq N$. To construct an updated closed testing procedure there are three forms of null hypoothesis to consider.\\

\textbf{$H_e$:} Let $\alpha'_e$ be the conditional error rate for the test of $H_e$ at the time the $N$ hypotheses are added, the test of $H_e$ requires the probability of falsely rejecting $H_e$ does not exceed $\alpha'_e$.\\

\textbf{$H_n$:} With no existing test for $H_n$ it must be tested at level $\alpha$.\\

\textbf{$H_e \cap H_n$:} $H_e \cap H_n \implies H_e$ and hence the data already available for $H_e$ is distributed such that computing the corresponding conditional error $\alpha'_e$ will ensure that an equation of the form~\ref{eq:con} holds. Thus we incorporate the existing information into the test of $H_e \cap H_n$ by constructing it such that the probability of falsely rejecting $H_e \cap H_n$ does not exceed $\alpha'_e$.\\

Any intersection of the form $H_e \cap H_n$ must be constructed in this way. While proposing changes to the trial one may or may not add a new hypotheses $H_n$: in the case where $H_n$ added then $H_e \cap H_n$ may be based on the data relating to both $H_e$ and $H_n$ and is tested at $\alpha'_e$; while if $H_n$ is not added the test for $H_e \cap H_n$ is implicitly that of $H_e$ also tested at $\alpha'_e$. In either case the test of $H_e \cap H_n$ is tested at $\alpha'_e$ ensuring that an equation of the form~\ref{eq:con} holds whatever decision is made while proposing changes to the trial design. Noting that any procedure that gives strong control of the FWER is a closed testing procedure \cite{burnett2021adaptive}. So we may add hypotheses to any procedure that ensures strong control of the FWER while maintaining the statistical integrity of the trial. The penalty for doing so compared is the test of hypotheses of the form $H_e \cap H_n$.

\section{Alteration of a Multi-Arm Multi Stage trial in progress}

\subsection{Multi-arm multi-stage trials}
\label{sec:mamsrec}

With multiple experimental treatments to compare with the control, we should consider a MAMS design \citep{jaki2015multi,wason2016some}. This allows us to compare the treatments in the same trial, while incorporating pre-planned interim analyses to facilitate early stopping. This ensures: poorly performing treatments may be dropped for futility; alternatively the trial may be stopped early to declare efficacy, reducing the overall development time and number of patients. This early stopping can be done while formally testing null hypotheses and controlling the FWER (Equation~\ref{eq:FWER}), through the use of generalised Dunnett testing procedures \citep{magirr2012generalized}. We use the extension of this proposed by \cite{urach2016multi}. This directly defines all elements of the closed test allowing us to directly apply our rule for adding hypotheses from Section~\ref{sec:gen}.

Suppose we have $K$ novel treatments, $T_1,...,T_K$ to compare against a common control. We define the null hypotheses $H_{0i}:\theta_i \leq 0$ and corresponding alternatives $H_{1i}:\theta_i > 0$ for all $i=1,...,K$. A MAMS designs will simultaneously test these $K$ hypotheses over $J$ analyses.\\

Let $n$ be the number of patients to be recruited to the control arm in the first stage of the trial. We assume patients are randomised at the desired rate in each stage of the trial. At analysis $j=1,...,J$ the trial will have recruited $r_{k}^{(j)}n$ patients to treatment $k=0,1,...,K$ ($r_0^{(1)}=1$ by construction). Treatments may be dropped futility at each analysis (and removed from any further consideration), suppose treatment $k*$ is stopped at analysis $j*$ we have $r_{k*}^{(j)}=r_{k*}^{(j*)}$ for all $j \geq j*$. If all $T_1,...,T_K$ are dropped for futility the trial stops recruiting. Alternatively the trial may stop early if a treatment or treatments have been selected for further study, such as when the trial is stopped due to a treatment-control comparison yielding statistical significance \citep{urach2016multi}.\\

From the observations at each stage $j=1,...,J$ and treatment $k=1,...,K$ we construct estimates $\hat{\theta}_{k}^{(j)}$. Then defining
\begin{equation*}
\mathcal{I}_{k}^{(j)} = \frac{r_{k}^{(j)}r_{0}^{(j)}n}{\sigma^2(r_{k}^{(j)}+r_{0}^{(j)})},
\end{equation*}
we find the corresponding Z-values
\begin{equation*}
Z_{k}^{(j)} = \hat{\theta}_{k}^{(j)}\sqrt{\mathcal{I}_{k}^{(j)}}.
\end{equation*}\\

In the testing procedures that follow we require that the ratio of patients assigned to each treatment remains consistent through each stage of the trial, that is
\begin{equation*}
\frac{r_0^{(j)}}{r_k^{(j)}} = \frac{r_0^{(l)}}{r_k^{(l)}}
\end{equation*}
for all $k = 1,...,K$ and $j,l = 1,...,J$ \citep{koenig2008adaptive}.

\subsection{The Generalised Dunnett procedure}
\label{sec:genDun}

Recall that $R$ is the event that we reject one or more true null hypothesis then extending Equation~\ref{eq:FWER} to $K$ null hypothesis strong control requires that 
\begin{equation*}
\mathbb{P}_{\boldsymbol{\theta}}(R) \leq \alpha \text{ for all } \boldsymbol{\theta}=(\theta_1,...,\theta_K).
\label{eq:FWER2}
\end{equation*} 
The generalised Dunnett method \citep{magirr2012generalized} simultaneously tests the null hypotheses, defining group sequential testing boundaries that account for the correlation structure of comparing multiple treatments to control to achieve the desired FWER.\\

We define efficacy boundaries $\boldsymbol{u} = (u_1,...,u_J)$ where the null hypothesis in treatment group $k=1,...,K$, $H_{0k}$, is rejected at analysis $j$ if $Z_{k}^{(j)} > u_j$ (and the trial is stopped). We define futility stopping boundaries $\boldsymbol{l} = (l_1,...,l_J)$ where if $Z_{k}^{(j)} < l_j$ the corresponding treatment is dropped for futility.\\

To achieve strong control of the FWER it is sufficient to choose $\boldsymbol{u}$ and $\boldsymbol{l}$ such that under the global null, $\theta_1=...=\theta_K=0$ which we denote by $\boldsymbol{0}$, 
\begin{equation*}
\mathbbm{P}_{\boldsymbol{0}}(R) \leq \alpha
\label{eq:gFWER}
\end{equation*}
\citep{magirr2012generalized}. Such testing boundaries may be computed using familiar group sequential theory testing.

\subsection{Group sequential closed testing}
\label{sec:gsc}

Let $\mathcal{K}$ be the set such for any $I \subseteq (1,...,K)$ we have that $\cap_{i\in I} H_{0i} \in \mathcal{K}$. Constructing tests for each $H_{0m} \in \mathcal{K}$ at level $\alpha$. We reject $H_{0k}$ globally when all tests including $H_{0k}$ are rejected at level $\alpha$ for $k=1,...,K$.\\

The generalised Dunnett defines the test for the intersection of all null hypotheses $H_{01}\cap ...\cap H_{0K}$ and implicitly tests all $H_{0m} \in \mathcal{K}$ using the same $\boldsymbol{u}$ and $\boldsymbol{l}$. \cite{urach2016multi} extend this by directly defining all tests required for the closed testing procedure; for each $H_{0m} \in \mathcal{K}$ testing boundaries $\boldsymbol{u}_m = (u_{1,m},...,u_{J,m})$ are in Section~\ref{sec:genDun}, $H_{0m}$ is rejected at stage $j$ if $Z_{k}^{(j)} > u_{j,m}$. The futility boundaries $\boldsymbol{l} = (l_1,...,l_J)$ must be the same for all hypotheses.\\

\subsection{Adding experimental treatment arms}

Suppose at the $J'^{th}$ ($J' \in (1,...,J)$) interim analysis of a MAMS trial in progress we wish to add $T \geq 1$ new treatments. We now have up to $K'= K+1+T$ treatments in total (in the case that all $K+1$ original treatment arms are all still in the trial). We have planned recruitment $r_{k}^{(j)}n$ for treatment $k=1,...,K+T$ at stage $j=1,...,J$ where $r_k^{(j)}=0$ for all $k > K$. When modifying the trial we define a modified recruitment plan, recruiting $r_{k}^{\prime(j)}n$ patients for each treatment $k = 1,...,K+T$ at each remaining stage of the trial  $j=J',...,J$ (we could also use this opportunity to modify the number of stages); for $j \leq J'$ we know that $r_{k}^{\prime(j)}n = r_{k}^{(j)}n$ while for $j > J'$ we fix the planned recruitment for the remainder of the trial at this point. As in Section~\ref{sec:simpadd} we use the independent increments of the Z-values splitting the trial according to patients recruited before and after the $J'^{th}$ analysis. For $j = J'+1 ,..., J$ and $k = 0,1,...,K$ the sample that would have been recruited is given by $r_{k}^{\star(j)} = r_{k}^{(j)} - r_{k}^{(J')}$, from which we compute Z-values $Z_{k}^{\star(j)}$. For each $k=1,...,K$ and $j=J'+1,...,J$ we define weights,
$$w_{1,k}^{(j)} = \sqrt{\frac{r_{k}^{(J')} + r_{0}^{(J')}}{r_{k}^{(j)} + r_{0}^{(j)}}},$$
$$w_{2,k}^{(j)} = \sqrt{1- w_{1,k}^{(j)}w_{1,k}^{(j)}}$$
and re-construct the Z-values for the remainder of the trial as 
$$Z_{k}^{(j)} = w_{1,k}^{(j)} Z_{k}^{(J')} + w_{2,k}^{(j)}  Z_{k}^{\star(j)}.$$
Weighting together the Z-values in this way will allow us to modify the ratio of patients recruited to each treatment at the time of the design modification. As in Section~\ref{sec:mamsrec} these ratios must remain fixed for all stages of the trial after the modification has been made. 

\subsection{Incorporating additional hypotheses}
\label{sec:tests}

We now have null hypotheses $H_{0i}:\theta_i \leq 0$ and alternatives $H_{1i}:\theta_i > 0$ for all $i=1,...,K+T$, and require strong control of the FWER across all $K+T$ tests. We construct a closed testing procedure following the rule introduced in Section~\ref{sec:gen}.  We define three sets: the set of existing null hypotheses $H_{01},...,H_{0K}$ and all intersections, $\mathcal{K}$; the set of added null hypotheses $H_{0K+1},...,H_{0K+T}$  and all intersections, $\mathcal{T}$; and the set of all intersections between existing and added null hypotheses, $\mathcal{KT}$.\\

The conditional error rate of each test for $H_{0m} \in \mathcal{K}$ is maximised under the global null \citep{stallard2015flexible}. Given the existing estimates, $\hat{\boldsymbol{\theta}}^{(J')} = (\hat{\theta}_{1}^{(J')},...,\hat{\theta}_{k}^{(J')})$ and under the originally planned trial described in Sections~\ref{sec:mamsrec}, \ref{sec:genDun} and \ref{sec:gsc} we write the conditional error for each $H_{0m} \in \mathcal{K}$ under the global null as 
\begin{equation*} 
B_{m}(\hat{\boldsymbol{\theta}}^{(J')}) = \mathbb{P}_{\boldsymbol{0}}(\text{Reject }H_{0m}|\hat{\boldsymbol{\theta}}^{(J')}) \leq \alpha.
\end{equation*}
As in Equation~\ref{eq:con} we have that under the global null 
\begin{equation*}
\int_{\hat{\boldsymbol{\theta}}^{(J')}}f(\hat{\boldsymbol{\theta}}^{(J')})B_{m}(\hat{\boldsymbol{\theta}}^{(J')}) \mathrm{d}\hat{\boldsymbol{\theta}}^{(J')} = \alpha
\label{eq:condint}
\end{equation*}
as required. It is useful to re-write the testing boundaries for each $H_{0m} \in \mathcal{K}$ in terms of only the data collected after stage $J'$, that is for $j=J'+1,...,J$ and $k=1,...,K$
$$u_{j,k,m} = \frac{u_{j,m} -w_{1,k}^{(j)} Z_{k}^{(J')}}{w_{2,k}^{(j)}}$$
for $k=1,...,K$, $H_{0k}$ is rejected at stage $j$ of the trial if $Z_{k}^{\star(j)} > u_{j,k,m}$ and
$$l_{k,j,m} = \frac{l_{j,m} -w_{1,k}^{(j)} Z_{k}^{(J')}}{w_{2,k}^{(j)}}$$
where if $Z_{k}^{\star(j)} < l_{k,j,m}$ $T_k$ is dropped for futility. This allows computation of the conditional error rate based $Z_{k}^{\star(j)}$ for $j = J'+1 ,..., J$ and $k = 0,1,...,K$.\\

For  each $H_{0m} \in \mathcal{K}$ the hypothesis test must be constructed at level $B_{m}(\hat{\boldsymbol{\theta}}^{(J')})$. For each $H_{0m} \in \mathcal{T}$ the hypothesis test must be constructed at level $\alpha$. For each $H_{0m} \in \mathcal{KT}$ the hypothesis test must be constructed at level $B_{m}(\hat{\boldsymbol{\theta}}^{(J')})$. This ensures each test for the trail as a whole is constructed at level $\alpha$ as required, while including any existing trial data and allowing for any changes to recruitment for $T_0,T_1,...,T_K$, and the FWER is strongly controlled.\\

For each hypothesis $H_{0m} \in K \cup T \cup KT$ we define the testing boundaries for the modified trial at the required error rate $\boldsymbol{u}'_m = (u'_{J'+1,m},...,u'_{J,m})$ and  $\boldsymbol{l}'_m = (l'_{J'+1,m},...,l'_{J,m})$. At stage $j = J'+1 ,..., J$ for treatment $k = 0,1,...,K$ the recruitment is governed by $r_{k}^{\star\prime(j)} = r_{k}^{\prime(j)} - r_{k}^{(J')}$, with corresponding Z-values $Z_{k}^{\star\prime(j)}$. For each experimental treatment from the first stage of the trial $k=1,...,K$ we define weights for data before and after stage $J'$, for $j=J'+1,...,J$ and $k=1,...,K$
$$w_{1,k}^{\prime(j)} = \sqrt{\frac{r_{k}^{(J')} + r_{0}^{(J')}}{r_{k}^{\prime(j)} + r_{0}^{\prime(j)}}},$$
$$w_{2,k}^{\prime(j)} = \sqrt{1- w_{1,k}^{\prime(j)}w_{1,k}^{\prime(j)}}$$
and construct the Z-values for for the hypothesis tests as
$$Z_{k}^{\prime(j)} = w_{1,k}^{\prime(j)} Z_{k}^{(J')} + w_{2,k}^{\prime(j)}  Z_{k}^{\star\prime(j)}$$
allowing us to write the testing boundaries for each $H_{0m} \in \mathcal{K}$ in terms of only the data collected after stage $J'$, that is for $j=J'+1,...,J$ and $k=1,...,K$
$$u'_{j,k,m} = \frac{u'_{j,m} -w_{1,k}^{\prime(j)} Z_{k}^{(J')}}{w_{2,k}^{\prime(j)}}$$
rejecting $H_{0m} \in \mathcal{K}$ at analysis $j=J'+1,...,J$ if $Z_{k}^{\star\prime(j)} > u_{j,k,m}$ and
$$l'_{k,j} = \frac{l'_{j} -w_{1,k}^{\prime(j)} Z_{k}^{(J')}}{w_{2,k}^{\prime(j)}}$$
where if $Z_{k}^{\star\prime(j)} < l_{k,j}$ $T_k$ dropped for futility (note for $k>K$ $u'_{j,m}$ and $l'_{j}$). With this in place $\boldsymbol{u}'_m $ and $\boldsymbol{l}'_m$ may be computed as per the generalised Dunnett test.

\section{Example}
\label{sec:exam}
\subsection{An illustrative example}

For the initial design consider a three stage trial to compare two treatments with a control, recruiting $n=10$ patients to each treatment at each stage of the trial; that is $J=3$, $K=2$ and $\boldsymbol{r}_k=(1,2,3)$ for $k=0,1,2$. Under this design we test the null hypotheses $H_{01}:\theta_1 \leq 0$ and $H_{02}:\theta_2 \leq 0$. The testing boundaries are constructed for a FWER of $\alpha = 0.05$, let $\delta = \Phi^{-1}(0.75)\sqrt{2}$  and $\sigma = 1$. At a configuration of $\boldsymbol{\theta} = (\delta,0)$ we have a target power of $1-\beta = 0.9$. Defining the triangular testing boundaries \citep{whitehead1997design} we first compute the testing boundary for $H_{01}\cap H_{02}$ using the \texttt{mams()} function of the MAMS package in R \citep{jaki2019r}. This sets futility boundary for all tests with the upper boundaries computed for testing both $H_{01}$ and $H_{02}$ separately.\\

Suppose after the first analysis $J'=1$ we add two further treatments $T=2$, adding the null hypotheses $H_{03}:\theta_3 \leq 0$ and $H_{04}:\theta_4 \leq 0$. Given $\boldsymbol{Z}^{(1)} = (2,1.5)$, the trial would continue in all arms at the interim analysis. Computing the conditional error rate for each existing test we construct all required tests as described in Section~\ref{sec:tests}. Using triangular testing boundaries ensuring all lower boundaries correspond to those of $H_{01}\cap H_{02}\cap H_{03}\cap H_{04}$. We continue recruiting 10 patients per treatment per stage, allowing for a maximum total sample size of 130 patients.\\

Table~\ref{tab:condsim} shows the operating characteristics of the updated trial based on 1,000,000 simulations of the remainder of the trial. Due to the tests being conditional on the first stage observations the probabilities of rejecting the null hypotheses under the global null are not 0.05. Since $Z_1^{(1)} > Z_2^{(1)}$ we observe higher probabilities of rejecting $H_{01}$ than $H_{02}$ for equivalent values of $\theta_1$ and $\theta_2$, for example when $\theta_1 = \theta_2 = \delta$ the probability of rejecting $H_{01}$ is 0.13 higher. Similarly since $Z_1^{(1)} >0$ and $Z_2^{(1)} >0$ the probability of rejecting $H_{01}$ or $H_{02}$ is higher than the probability of rejecting $H_{03}$ or $H_{04}$, for example when $\theta_1 = \theta_2 =\theta_3 = \theta_4 = \delta$ we have probabilities of 0.94, 0.81, 0.59 and 0.59 of rejecting $H_{01}$,$H_{02}$, $H_{03}$ and  $H_{04}$ respectively. We also see the benefit of incorporating all treatments in the same trial, with a reduction in the expected sample size and a chance to reject multiple null hypotheses, when there are more beneficial treatments overall.

\begin{table}
\makebox[\linewidth]{
\begin{tabular}{c | c c c c c}
$\boldsymbol{\theta}$ & $\mathbb{P}_{\boldsymbol{\theta}}(R_1)$ & $\mathbb{P}_{\boldsymbol{\theta}}(R_2)$ & $\mathbb{P}_{\boldsymbol{\theta}}(R_3)$ & $\mathbb{P}_{\boldsymbol{\theta}}(R_4)$ & $\mathbb{E}_{\boldsymbol{\theta}}(N)$\\\hline
$( 0 , 0 , 0 , 0 )$  										 &  0.19  &  0.09  &  0.02  &  0.02  &  72 \\
$( \delta , 0 , 0 , 0 )$  							 &  0.97  &  0.08  &  0.02  &  0.02  &  54 \\
$( 0 , \delta , 0 , 0 )$  							 &  0.23  &  0.94  &  0.02  &  0.02  &  59 \\
$( \delta , \delta , 0 , 0 )$  					 &  0.96  &  0.83  &  0.02  &  0.02  &  53 \\
$( 0 , 0 , \delta , 0 )$  							 &  0.16  &  0.07  &  0.79  &  0.02  &  66 \\
$( \delta , 0 , \delta , 0 )$  					 &  0.95  &  0.08  &  0.57  &  0.02  &  54 \\
$( 0 , \delta , \delta , 0 )$  					 &  0.22  &  0.88  &  0.63  &  0.03  &  58 \\
$( \delta , \delta , \delta , 0 )$  		 &  0.95  &  0.82  &  0.55  &  0.03  &  53 \\
$( 0 , 0 , \delta , \delta )$  					 &  0.15  &  0.07  &  0.69  &  0.68  &  63 \\
$( \delta , 0 , \delta , \delta )$  		 &  0.93  &  0.08  &  0.57  &  0.58  &  54 \\
$( 0 , \delta , \delta , \delta )$  		 &  0.21  &  0.84  &  0.63  &  0.63  &  58 \\
$( \delta , \delta , \delta , \delta )$  &  0.94  &  0.81  &  0.59  &  0.59  &  53 \\
\multicolumn{6}{c}{}\\
\end{tabular}}
\makebox[\linewidth]{
\begin{tabular}{c | c c c c c}
$\boldsymbol{\theta}$  & Fail to reject  & Reject one & Reject two & Reject three & Reject four\\\hline
$( 0 , 0 , 0 , 0 )$  										 &  0.76  &  0.17  &  0.06  &  0.01  &  0.00 \\
$( \delta , 0 , 0 , 0 )$  							 &  0.03  &  0.87  &  0.09  &  0.01  &  0.00 \\
$( 0 , \delta , 0 , 0 )$  							 &  0.06  &  0.71  &  0.21  &  0.02  &  0.00 \\
$( \delta , \delta , 0 , 0 )$  					 &  0.01  &  0.19  &  0.77  &  0.03  &  0.01 \\
$( 0 , 0 , \delta , 0 )$  							 &  0.18  &  0.65  &  0.11  &  0.04  &  0.01 \\
$( \delta , 0 , \delta , 0 )$  					 &  0.02  &  0.44  &  0.47  &  0.07  &  0.01 \\
$( 0 , \delta , \delta , 0 )$  					 &  0.03  &  0.38  &  0.41  &  0.16  &  0.02 \\
$( \delta , \delta , \delta , 0 )$  		 &  0.01  &  0.15  &  0.37  &  0.45  &  0.03 \\
$( 0 , 0 , \delta , \delta )$  					 &  0.08  &  0.43  &  0.36  &  0.09  &  0.04 \\
$( \delta , 0 , \delta , \delta )$ 		   &  0.01  &  0.28  &  0.30  &  0.33  &  0.07 \\
$( 0 , \delta , \delta , \delta )$  		 &  0.02  &  0.26  &  0.25  &  0.30  &  0.16 \\
$( \delta , \delta , \delta , \delta )$  &  0.01  &  0.12  &  0.24  &  0.22  &  0.42 
\end{tabular}}

\caption{Operating characteristics for the remainder of the trial given $\boldsymbol{Z}_1 = (2,1.5)$ under corresponding configuration $\boldsymbol{\theta}$. Where $R_i$ is the event that $H_{0i}$ is rejected and $N$ is the total sample size (note 30 participants already recruited).}
\label{tab:condsim}
\end{table}

\subsection{Comparison of performance}

We compare our proposed method with two options that maintain the integrity of the results given that observations are already available from the trial: option one is to conduct a separate MAMS trial comparing the new treatments with the control in addition to the trial already in progress; option two is to conclude the current trial and start a new trial incorporating all four experimental treatments. In examining these unmodified designs we make no use of the previous trial data, meaning these trials do not benefit from the patients already recruited.\\

As before we add two treatments, $T=2$, at the first analysis, $J'=1$. We keep all other parameters as before, keeping them consistent for each design. We estimate the operating characteristics of our proposed method based on 10,000 simulations. This is a lower number of simulations than would be ideal, due to the computationally intensive nature of the simulation. In practice we do not expect this to be used as a pre-planned scheme and hence only one set of updated testing boundaries need computing, making longer simulations more viable as seen in Section~\ref{sec:exam}. For the unmodified option one, we use the original trial for $H_{01}$ and $H_{02}$ and compute boundaries for a two stage trial for $H_{03}$ and $H_{04}$. For the unmodified option two, we compute boundaries for a two stage trial for $H_{01}$, $H_{02}$, $H_{03}$ and $H_{04}$. We use 1,000,000 simulations to estimate the operating characteristics for each unmodified design. \\

We break the operating characteristics of our proposed method down, with Table~\ref{tab:stop} showing the behaviour of the first stage of the trial and Table~\ref{tab:prop1} showing trial that continues beyond the first interim analysis. We see a relatively high probability of the trial concluding at the first analysis, before the treatments are added; this shows the first stage data should not be disregarded. If the trial continues beyond the first stage we observe a similar pattern to that shown in Table~\ref{tab:condsim}. With lower probabilities of rejecting $H_{03}$ or $H_{04}$ than rejecting $H_{01}$ or $H_{02}$. The probabilities of rejection are lower than Table~\ref{tab:condsim} since $l^{(1)} = 0$ allows for less promising stage-1 Z-values to progress the trial beyond the first analysis.\\

\begin{table}[ht]
\makebox[\linewidth]{
\begin{tabular}{c | c  c c}
$\boldsymbol{\theta}$ & $\mathbb{P}_{\boldsymbol{\theta}}(\text{Continue beyond stage 1})$ & $\mathbb{P}_{\boldsymbol{\theta}}(\text{Stop for efficacy at first analysis})$ & $\mathbb{P}_{\boldsymbol{\theta}}(\text{Stop for futility at first analysis})$ \\\hline
$( 0 , 0 )$  					 &  0.64 & 0.02 & 0.34 \\
$( \delta , 0 )$  		 &  0.64 & 0.35 & 0.01 \\
$( \delta , \delta )$  &  0.46 & 0.53 & 0.00 \\
\end{tabular}}
\caption{Performance of the original trial at the first interim analysis.}
\label{tab:stop}
\end{table}

\begin{table}
\makebox[\linewidth]{
\begin{tabular}{c | c c c c c}
$\boldsymbol{\theta}$ & $\mathbb{P}_{\boldsymbol{\theta}}(R_1)$ & $\mathbb{P}_{\boldsymbol{\theta}}(R_2)$ & $\mathbb{P}_{\boldsymbol{\theta}}(R_3)$ & $\mathbb{P}_{\boldsymbol{\theta}}(R_4)$ & $\mathbb{E}_{\boldsymbol{\theta}}(N)$\\\hline
$(0,0,0,0)$  										&  0.03  &  0.02  &  0.01  &  0.01  &  78  \\
$(\delta,0,0,0)$  							&  0.91  &  0.00  &  0.01  &  0.01  &  71  \\
$(\delta,\delta,0,0)$  					&  0.79  &  0.84  &  0.02  &  0.02  &  77  \\
$(0,0,\delta,0)$  							&  0.01  &  0.02  &  0.71  &  0.01  &  78  \\
$(\delta,0,\delta,0)$  					&  0.86  &  0.01  &  0.64  &  0.03  &  71  \\
$(\delta,\delta,\delta,0)$  		&  0.82  &  0.82  &  0.61  &  0.01  &  75  \\
$(0,0,\delta,\delta)$  					&  0.02  &  0.02  &  0.63  &  0.63  &  79  \\
$(\delta,0,\delta,\delta)$  		&  0.82  &  0.01  &  0.60  &  0.60  &  70  \\
$(\delta,\delta,\delta,\delta)$ &  0.77  &  0.77  &  0.61  &  0.61  &  76  \\
\multicolumn{6}{c}{ }
\end{tabular}}
\makebox[\linewidth]{
\begin{tabular}{c | c c c c c}
$\boldsymbol{\theta}$  & Fail to reject  & Reject one & Reject two & Reject three & Reject four\\\hline
$(0,0,0,0)$  										&  0.94  &  0.05  &  0.01  &  0.00  &  0.00  \\
$(\delta,0,0,0)$  							&  0.08  &  0.89  &  0.02  &  0.00  &  0.00  \\
$(\delta,\delta,0,0)$  					&  0.04  &  0.30  &  0.62  &  0.04  &  0.00  \\
$(0,0,\delta,0)$  							&  0.29  &  0.68  &  0.03  &  0.00  &  0.00  \\
$(\delta,0,\delta,0)$  					&  0.05  &  0.40  &  0.52  &  0.03  &  0.00  \\
$(\delta,\delta,\delta,0)$  		&  0.01  &  0.19  &  0.35  &  0.44  &  0.01  \\
$(0,0,\delta,\delta)$  					&  0.12  &  0.48  &  0.37  &  0.02  &  0.00  \\
$(\delta,0,\delta,\delta)$  		&  0.03  &  0.27  &  0.35  &  0.33  &  0.01  \\
$(\delta,\delta,\delta,\delta)$ &  0.02  &  0.15  &  0.21  &  0.30  &  0.32  
\end{tabular}
}
\caption{Under our proposed update procedure, probabilities of rejecting null hypotheses and expected sample size under the corresponding configuration of $\boldsymbol{\theta}$ for our proposed update procedure when the trial continues beyond the first stage. Where $R_i$ is the event that we reject $H_{oi}$ and $N$ is the total sample size (including the 30 patients  included in stage one).}
\label{tab:prop1}
\end{table}

Comparing our proposed procedure with option one shown in Table~\ref{tab:opt1}, conducting two separate trials produces similar probabilities for rejecting $H_{01}$ or $H_{02}$. Our method is sensitive to $\theta_3$ and $\theta_4$ due to their ability to also conclude the trial early. Two separate trials increase the probabilities of rejecting $H_{03}$ or $H_{04}$; this is partially due to the disconnect between trials, if one concludes early the other may continue and reject and null hypothesis. Given this and that patients are recruited to the control in both trials we see that our proposed method significantly reduces the expected sample size, with 70--80 patients including the first stage of the trial for trials that continue beyond the first stage (for the trial as a whole this expected sample size drops to 50--60 over the scenarios we have examined)  whereas option one requires 90--95 patients. There are two key flaws in option one: while this method incorporates all existing data for $H_{01}$ and $H_{02}$ there is no multiplicity adjustment between the existing and added hypotheses, as we have two separate trials of two null hypotheses each with a FWER of $\alpha$; if we wish to select some subset of treatments for further study, there is no guarantee of direct comparability between each trial.\\

\begin{table}
\makebox[\linewidth]{
\begin{tabular}{c | c c c c c c}
$\boldsymbol{\theta}$ & $\mathbb{P}_{\boldsymbol{\theta}}(R_1)$ & $\mathbb{P}_{\boldsymbol{\theta}}(R_2)$ & $\mathbb{E}_{\boldsymbol{\theta}}(N_1)$ & $\mathbb{P}_{\boldsymbol{\theta}}(R_3)$ & $\mathbb{P}_{\boldsymbol{\theta}}(R_4)$ & $\mathbb{E}_{\boldsymbol{\theta}}(N_2)$\\\hline
$( 0 , 0 , 0 , 0  )$  									 &  0.03  &  0.03  &  49  &  0.03  &  0.03  &  38 \\
$( \delta , 0 , \delta , 0 )$  					 &  0.93  &  0.02  &  47  &  0.82  &  0.04  &  39 \\
$( \delta , \delta , \delta , \delta )$  &  0.81  &  0.81  &  45  &  0.77  &  0.77  &  39 \\
\multicolumn{7}{c}{ }
\end{tabular}}
\makebox[\linewidth]{
\begin{tabular}{c | c c c c c c}
 & \multicolumn{3}{c}{Original trial} & \multicolumn{3}{c}{Additional trial}\\
$\boldsymbol{\theta}$  & Fail to reject  & Reject one & Reject two &  Fail to reject  & Reject one & Reject two\\\hline
$( 0 , 0 , 0 , 0  )$  									 &  0.95 & 0.04 & 0.01  & 0.95 & 0.04 & 0.01 \\
$( \delta , 0 , \delta , 0 )$  					 &  0.07 & 0.90 & 0.02  & 0.18 & 0.78 & 0.04 \\
$( \delta , \delta , \delta , \delta )$  &  0.02 & 0.34 & 0.64  & 0.07 & 0.31 & 0.62
\end{tabular}
}
\caption{Under two separate trials, probabilities of rejecting null hypotheses and expected sample size under the corresponding configuration of $\boldsymbol{\theta}$ for our option one assuming the trial continues beyond the interim analysis. Where $R_i$ is the event that we reject $H_{oi}$, $N_1$ is the total sample size in the original trial and $N_2$ is the total sample size in the additional trial.}
\label{tab:opt1}
\end{table}

Comparing the operating characteristics of option two in Table~\ref{tab:opt2} with our method in Table~\ref{tab:prop1}, we see that the probabilities of rejecting $H_{01}$ or $H_{02}$ are lower while the probabilities of rejecting $H_{03}$ or $H_{04}$ are similar, this  leads to a reduction in the probabilities of rejecting multiple hypotheses. For example, when $\theta_1 = \theta_2 =\theta_3 = \theta_4 = \delta$ the probabilities of rejecting $H_{01}$,$H_{02}$, $H_{03}$ and $H_{04}$ are 0.77, 0.77,  0.61  and  0.61 respectively, while they are all 0.64 under option two and the probability of rejecting two or more hypotheses falls by 0.12 compared to our proposed method. The expected sample size of the trial conducted under option two is reduced by 8--15 patients this does not account for the fact that 30 patients have been recruited who do not contribute to the result.

\begin{table}
\makebox[\linewidth]{
\begin{tabular}{c | c c c c c}
$\boldsymbol{\theta}$ & $\mathbb{P}_{\boldsymbol{\theta}}(R_1)$ & $\mathbb{P}_{\boldsymbol{\theta}}(R_2)$ & $\mathbb{P}_{\boldsymbol{\theta}}(R_3)$ & $\mathbb{P}_{\boldsymbol{\theta}}(R_4)$ & $\mathbb{E}_{\boldsymbol{\theta}}(N)$\\\hline
$(0,0,0,0)$  										&  0.02  &  0.02  &  0.02  &  0.02  &  62 (+30)\\
$(\delta,0,0,0)$  							&  0.75  &  0.01  &  0.01  &  0.01  &  63 (+30)\\
$(\delta,\delta,0,0)$  					&  0.67  &  0.67  &  0.02  &  0.02  &  62 (+30)\\
$(0,0,\delta,0)$  							&  0.01  &  0.01  &  0.75  &  0.01  &  63 (+30)\\
$(\delta,0,\delta,0)$  					&  0.67  &  0.02  &  0.67  &  0.02  &  62 (+30)\\
$(\delta,\delta,\delta,0)$  		&  0.64  &  0.64  &  0.64  &  0.03  &  62 (+30)\\
$(0,0,\delta,\delta)$  					&  0.02  &  0.02  &  0.67  &  0.67  &  62 (+30)\\
$(\delta,0,\delta,\delta)$  		&  0.64  &  0.03  &  0.64  &  0.64  &  62 (+30)\\
$(\delta,\delta,\delta,\delta)$ &  0.64  &  0.64  &  0.64  &  0.64  &  63 (+30)\\
\multicolumn{6}{c}{ }
\end{tabular}}
\makebox[\linewidth]{
\begin{tabular}{c | c c c c c}
$\boldsymbol{\theta}$  & Fail to reject  & Reject one & Reject two & Reject three & Reject four\\\hline
$(0,0,0,0)$  										&  0.95  &  0.04  &  0.01  &  0.00  &  0.00 \\
$(\delta,0,0,0)$  							&  0.24  &  0.73  &  0.02  &  0.00  &  0.00 \\
$(\delta,\delta,0,0)$  					&  0.11  &  0.43  &  0.43  &  0.02  &  0.01 \\
$(0,0,\delta,0)$  							&  0.24  &  0.73  &  0.02  &  0.00  &  0.00 \\
$(\delta,0,\delta,0)$  					&  0.11  &  0.43  &  0.43  &  0.02  &  0.00 \\
$(\delta,\delta,\delta,0)$  		&  0.07  &  0.31  &  0.28  &  0.33  &  0.03 \\
$(0,0,\delta,\delta)$  					&  0.11  &  0.43  &  0.43  &  0.02  &  0.00 \\
$(\delta,0,\delta,\delta)$  		&  0.07  &  0.31  &  0.27  &  0.33  &  0.03 \\
$(\delta,\delta,\delta,\delta)$ &  0.05  &  0.24  &  0.19  &  0.18  &  0.35  
\end{tabular}
}
\caption{Under starting a new trial incorporating all treatments, probabilities of rejecting null hypotheses and expected sample size under the corresponding configuration of $\boldsymbol{\theta}$ for option two assuming the trial continues beyond the interim analysis. Where $R_i$ is the event that we reject $H_{oi}$ and $N$ is the total sample size (note 30 additional patients are recruited but not used in the analysis).}
\label{tab:opt2}
\end{table}

\section{Discussion}

The motivation for adding a treatment to a trial in progress is clear. Should a new treatment become available it is desirable incorporate it allowing direct comparisons while preserving integrity and avoiding delays to the overall development process. Our proposed general framework for adding experimental treatments to a trial in progress builds upon the work of \cite{hommel2001adaptive}, allowing any trial with strong control of the FWER to add new hypotheses. This also allows other alterations to the design of the trial while ensuring that all information already collected is utilised in inference and decision making.\\

This framework can be applied in our motivational setting of MAMS platform trials \citep{meyer2021systematic,meyer2020evolution}. The examples in Section~\ref{sec:exam} demonstrate that this does indeed strongly control the FWER as expected. 

Our examples in Sections~\ref{sec:sim} and~\ref{sec:exam} show the penalty adding treatments in terms of the probability of rejecting the null hypotheses is marginal and only has a notable impact on the introduced arms, optimising the recruitment proportions across configurations of the true treatment effects may reduce the impact of this further. In addition the combination of utilising the existing data and the efficient use of control patients across the trial yields a reduction in the expected sample size when compared to alternatives that do not make such use of the existing data. The operating characteristics are not the primary motivation to adding treatments to a trial in progress. As for MAMS designs in general this allows for reduction in logistical and administrative effort and speeding up the overall development process as well as allowing direct comparisons of the treatments within the same trial.\\

The general framework for adding hypotheses to a trial in progress has broader application than, being applicable to any testing procedure that gives strong control of the FWER. The addition of hypotheses in this way allows for the incorporation of existing trial data into decisions about how to plan the remainder of the trial.

\section{Software}
Software relating to the examples in this paper is available at \href{https://github.com/Thomas-Burnett/Adding-treatments-to-clinical-trials-in-progress.git}{https://github.com/Thomas-Burnett/Adding-treatments-to-clinical-trials-in-progress.git}.

\section*{Acknowledgements}

This research was supported by the NIHR Cambridge Biomedical Research Centre (BRC-1215-20014). This report is independent research supported by the National Institute for Health Research (Prof Jaki's Senior Research Fellowship, NIHR-SRF-2015-08-001). T Jaki received funding from the UK Medical Research Council (MC\_UU\_00002/14). Franz K{\"o}nig is a member of the EU Patient-Centric Clinical Trial Platform (EU-PEARL) which has received funding from the Innovative Medicines Initiative 2 Joint Undertaking, grant No 853966. This Joint Undertaking receives support from the EU Horizon 2020 Research and Innovation Programme, EFPIA, Children's Tumor Foundation, Global Alliance for TB Drug Development, and SpringWorks Therapeutics. The views expressed in this publication are those of the authors. These views are not necessarily those of the NHS, the National Institute for Health Research or the Department of Health and Social Care (DHSC). The funders and associated partners are not responsible for any use that may be made of the information contained herein.

{
\setlength{\bibsep}{0.0pt}
\baselineskip20pt
\bibliographystyle{biorefs}

\bibliography{mams}
}

\end{document}